\documentclass[12pt]{iopart}
\usepackage{graphics,amssymb} 
\usepackage{graphicx}
\usepackage{color}
\usepackage {citesort}

\begin{document}

\title[]{Quench dynamics of neutral atoms in out-equilibrium one dimensional optical lattices}

\author{ Vicente Leyton$^{1,2}$, Arturo Arg\"uelles$^{1}$, Manuel Camargo$^{2}$}

\address{$^1$ Centro de Investigaci\'on en Ciencias B\'asicas, Universidad Santiago de Cali, Calle 5 No. 62 - 00 Cali, Colombia}

\address{$^2$ Centro de Investigaciones en Ciencias B\'asicas y Aplicadas, Universidad Antonio Nari\~no, Km 18 via Cali-Jamund\'i, 760030 Santiago de Cali, Colombia}

\ead{vicenley@gmail.com}

\begin{abstract}
A quantum simulator is proposed for nucleation and growth dynamics using an out-of equilibrium optical lattice. We calculate the density of neutral atoms in the lattice and we establish the connection with the Kolmogorov-Mehl-Johnson-Avrami model. Here we show that a Avrami equation can describe most of the evolution in time of the population growth in the lattice, coherence between neutral atoms leads a complex growth rate.  
\end{abstract}

\section{Introduction}


Ultracold bosonic and fermionic quantum gases are versatile and robust systems for probing fundamental condensed matter physics problems like quantum phase transitions\cite{Greiner2002,Bloch2008,Panas2017}, squeezed states in a Bose-Einstein condensate\cite{Orzel2001,Muessel2014},  Tonks-Girardeau gas in a one dimensional lattice\cite{Paredes2004,Wang2017}, vortices and superfluidity\cite{Zwierlein2005,Sensarma2006,Madeira2017}, as well as finding applications in quantum optics and quantum information processing like state selective production of molecules in optical lattices\cite{Rom2004,Yan2013} and induced oscillations between an atomic and molecular quantum gas.\cite{Mackie2005,Vitanov2017} Storing such ultracold quantum gases in artificial periodic potentials of light has opened innovative manipulation and control possibilities\cite{Bloch2005,Bloch2005a}, in many cases creating structures far beyond those currently achievable in typical condensed-matter physics. Ultracold quantum gases in optical lattices can in fact be considered as quantum simulators for the study of real materials\cite{Hague2017}, till the simulation of a Dirac field near an event horizon\cite{Rodriguez-Laguna2017}, due to they offers remarkably clean access to a particular hamiltonian and thereby serves as a model system for testing fundamental theoretical concepts.  

Storing interacting atoms in optical potentials is a very interesting setup.  Despite the presence of interaction between atoms, which lead to nonlinear terms in the Schr\"odinger-like equation (Gross-Pitaevskii equation),\cite{Pitaevskii2003} the macroscopic wavefunction still describes the quantum many-body system in the weakly interaction regime. If strong interacting regime with respect to the kinetic energy,  the system in general can not longer be described as a simple matter wave.  An example of the relevance of the interaction strength is the superfluid-to-Mott insulating state transition when the system go from a weakly interacting quantum system to a strongly correlated quantum many-body system system.\cite{Greiner2002}  For weak interactions the system form a Bose-Einstein condensate of matter.  The system tends to such state as the kinetic energy the kinetic energy is minimized for single particle wave function spread out throughout the lattice. For a strong inter-particle interaction relative to the kinetic energy, the system reaches the strongly correlated state or a Mott insulator state, in which the atoms are localized to single lattice sites. There the system cannot be described by giant coherent matter wave, and no interference pattern can be observed upon releasing the particle from the lattice. 

Mott insulators in one dimension have become attainable through the use of a deep two dimensional optical lattice loaded with a Bose-Einstein condensate (BEC).\cite{Stoeferle2004,Koehl2005,Koehl2004} In that setup the condensate splits up into several thousand individual one dimensional BECs in each of the potential tubes. Subsequently, a third lattice potential applied along the direction of the tubes can drive the transition to the Mott insulating state. Other possible mechanism to drive the system to the Mott insulating state is drive parametrically the tubes,\cite{Leyton2014} thus the on-site interaction can be tuned and a transition induced, by tuning the external modulation it can even mimic a Tonks-Girerdeau gas dynamics.  Experiments where a ultracold quantum gas has been controlled by means of periodic external modulation has been realized, in which the ballistic spreading of a localized Bose-Einstein condensate in the lowest band has been almost suppressed by the application of an time periodic external force in a one dimensional optical lattice.\cite{DRESE1997201}  In those experiments a one dimensional lattice is created in the tight-binding regime along the $x$ direction together with a tube-like harmonic confinement in transverse directions $y$ and $z$. In that setup a Bose-Einstein condensate is loaded (like ${}^{\rm 87}$Rb) in the lowest Bloch band of the lattice, localized in the center of the tube by additional trapping potential (along $x$). When this trap is removed the condensate expands along the tube, additionally a time periodic external force is applied shaking the lattice back and forth, for more details see \cite{Eckardt2017} and references therein.         

Recently the quench dynamics in Rydberg gases has been study in the classical and quantum regimes. There the Kolmorov-Johnson-Mehl-Avrami (KJMA) frame work was used for an analytical understanding of the quench dynamics with the coupling with a dissipative environment.\cite{Gribben2018} Following their approach we consider in this work the quench dynamics of neutral atoms in an one-dimensional system. We address to the nucleation problem when the lattice is build from a optical potential driven by an external field. 

In the following we shall introduce the time dependent Hamiltonian and the rotating wave approximations. In section 2 we shall introduce the out-equilibrium set up use to study KJMA nucleation process in optical lattices. In section 3 we describe the quench protocol for the nucleation and growth in the quantum regime. After that, in section 4 we analyze the population of the lattice in terms in the KJMA framework and we discuss the difference with the classical setup. We conclude in section 5. 

\section{Modulated one dimensional lattice}
We consider a one dimensional optical lattice along $x$ direction created by a laser beam with the optical potential $V_{\rm lattice} ({x}) = -V_0 \cos ( 2 \pi x / a)/2 + V_0/2$, with $a$ as the lattice size and $V_0 $ as the depth of the potential well. If the standing wave that constructs the optical lattice is generated by a directing a laser beam against a mirror, it is possible to modulate the position of the mirror periodically in time.\cite{DRESE1997201}  Thus, the optical potential become $V_{\rm mod} (x,t) = V_{\rm lattice } (x + A \cos (\omega t))$, with $A$ as the amplitude and $\omega$ the frequency of the mirror modulation. For weak driving strengths characterized by $A << a$ the modulated optical potential around the minimum $x_n = n  a$  reads as 
\begin{equation}
	V_{\rm mod}(x,t) \approx - \frac{V_0}{2} 
	   + F_0  \cos (\omega t) (x - x_n ) 
	   + \frac{\omega_{\rm lattice}}{2} (x - x_n)^2 \ , 
\end{equation}   
therein $F_0 = 2 \pi V_0 A/a $ and $\omega_{\rm lattice} = V_0 (2 \pi / a)^2$ are the effective driving strength and the onsite proper frequency respectively.  The resulting potential correspond to the simple case of a driven harmonic potential around each minimum $x_n$,  it is common in the study of laser-atom interactions in the dipole approximation.

In the lattice frame of reference the system can me modeled by the Hamiltonian 

\begin{eqnarray}\label{eq:model}
\tilde{H}(t) &=&  \omega_0 \sum_k  a^\dagger_k \, a^{}_{k}
 + F(t)  \sum_k (a_k^{\dagger } + a^{}_k)   
 + V \sum_k n_k n_{k+1} 
\nonumber \\  &&
 - J \sum_{k,k+1} (a^\dagger_k \, a^{}_{k+1} + {\rm h.c.})  
 + \frac{U}{2} \sum_k n_k (n_k -1)
\end{eqnarray} 
   
where $a_k$ ($a_k^\dagger$) destroys (creates) a particle on the $k$-th site, $n_k = a^\dagger_k a^{}_k$ the number operator, $\omega_0$ is the on-site energy,  $F(t) = f \cos (\omega t)$ is the external modulation under frequency $\omega$ and driving strength $f = F_0 x_0 /2$ with $x_0$ as the zero point fluctuations amplitude, $V$ the interaction energy of nearest neighbor bosons,  $J$ the hopping term that describes tunneling of particles between neighboring lattice sites, and $U$ the on-site interaction strength.
	
   It is convenient, for later purposes, switch to a frame of reference rotating under the angle $\theta(t) = \omega t$, by performing the canonical transformation $R(t) = \exp [- i \theta(t) \sum n_k ]$.  Here, we are interested on the quench dynamics of the lattice around twice its fundamental frequency, for $\omega \approx  \omega_0$. We further assume a weak modulation regime, where $F_0 x^2_0 / 2 << \omega_0$. This allow us to invoke the rotating wave approximation, in which the fast oscillating terms will be negligible around fundamental frequency for weak enough driving. By eliminating all fast oscillating terms from the transform Hamiltonian,  one obtains the following time-independent Hamiltonian

\begin{eqnarray}
	H &=& \delta \omega \, \sum_{k}  \, n_k + \frac{f}{2}  \, \sum_k (a^{\dagger}_k  + a_k^{}) + V \sum_k  n_k n_{k+1} 
	\nonumber \\ &&
	\nonumber \\  &&
	- J \sum_{k,k+1} (a^\dagger_k \, a^{}_{k+1} + {\rm h.c.})  
	+ \frac{U}{2} \sum_k n_k (n_k -1)   
\end{eqnarray}

There, we have introduced the detuning frequency $\delta \omega = \omega -  \omega_0$. In the following we shall introduce the quench protocol and calculate the density of particles in the lattice with the time-dependent-block-decimation method using open source software from Carr Theoretical research group.\cite{Wall2009}

\section{Quench dynamics}

As an initial preparation we consider an empty lattice $|00\cdots 0 \rangle $, in the regime under the condition $U >> J$ where the tunneling process is not favorable and the coming particles will fill the lattice.  In order to facilitate the filling process we set the detuning such that it cancels the interaction energy of adjacent filled sites, i.e. $\delta \omega  = -  V$, thus the Hamiltonian model become 
\begin{equation}
	H =  V \sum_k  (n_k-1) n_{k+1} +\frac{f}{2}  \, \sum_k (a^{\dagger}_k  + a_k^{}) 
\end{equation}
In the hardcore regime the onsite interaction term ($\propto U$) does not contribute to the dynamics. 
Opening a channel for the driving process opens also a channel for dissipation, this out-equilibrium dynamics is modeled by coupling the system to a bath composed by a infinite collection of harmonic oscillators named environment. For a weak system-environment coupling one can invoke the Born-Markov approximation leading to the Lindblad master equation\cite{Weiss2012}  
\begin{equation}
	\frac{d}{dt}\rho = - i [H, \rho] 
	  - \frac{\gamma}{2}\sum_k \left( a_k^\dagger a_k^{} \rho + \rho \, a_k^\dagger a_k^{} - 2 a_k^{}  \rho \, a_k^\dagger    \right) 
\end{equation}       
where the second term models decays and decoherence processes, with a rate $\gamma << \omega_0$. One can rewrite the above master master equation in an alternative way\cite{Daley2014}
\begin{equation}
	\frac{d}{dt} \rho =  - [H_{\rm eff} , \rho ] - \gamma \sum_k a_k \rho a^\dagger_k
\end{equation}
where $H_{\rm eff} =  H - i \gamma \sum_k n_k /2 $ as the effective Hamiltonian to model the dissipative dynamics. We can keep the effective Hamiltonian as a further approximation, and neglect the recycling part $- \gamma \sum_k  a_k \rho a^\dagger_k$.  From the external driving the system absorbs particles leading to the process $|00 \cdots 0 \rangle \rightarrow |0010\cdots 0 \rangle $, further by the interaction with the adjacent sites the system state can evolve as $ |0010 \cdots 0 \rangle \rightarrow |01110 \cdots 0 \rangle$. On the other hand, dissipation induces decay $|0010\cdots 0\rangle \rightarrow |0000 \cdots 0 \rangle$ and decoherence $(|0010 \cdots 0\rangle + |0000 \cdots 0\rangle)(|0001 \cdots 0\rangle + |0000 \cdots 0\rangle) \rightarrow |0000 \cdots 0\rangle + |0110 \cdots 0\rangle$. The lattice filling process is therefore analog to the nucleation and growth in classical setups like the KJMA framework.

\begin{figure}[h]
	\centering
	\includegraphics[width=\textwidth]{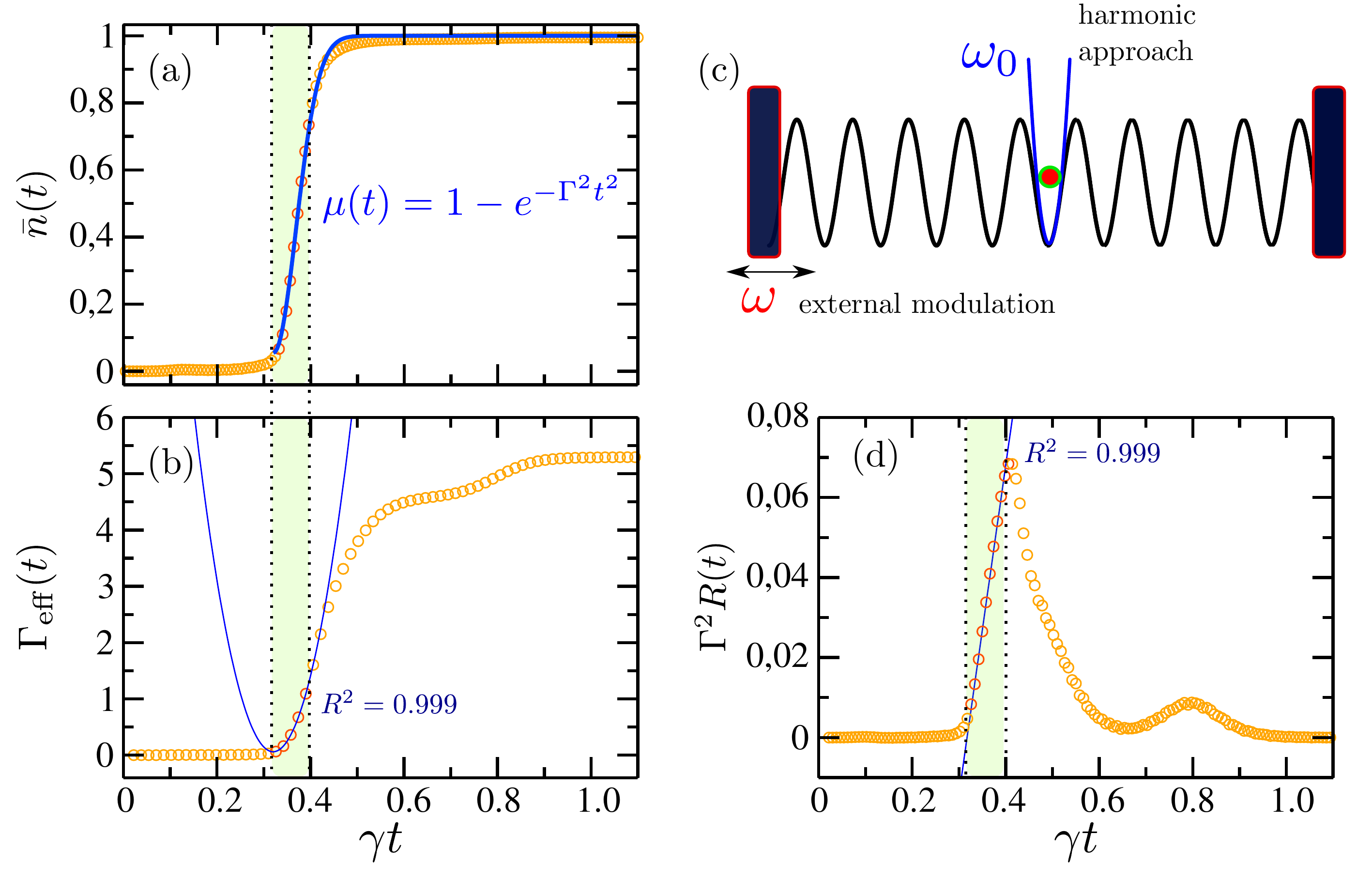}
	\caption{\label{fig:plot1} (a) Time evolution of the particle density for a lattice of $L = 200$ sites, interaction energy $V = 1$, driving strength $f = 10^{-1}V$, and damping rate $\gamma =  5 \times 10^{-2} V$. The time evolution of the effective nucleation rate is depicted in (b), where we have shadowed the region for $F_{\rm eff} (t) \propto t^2$, a quadratic fit of the points in that region is showed with a coefficient of determination $R^2 = 0.999$. In (c) we show a sketch of the modulated optical lattice in one dimension. In (d) we calculate the behavior of the growth function, a linear fit is show in the Fig.\ref{fig:plot1}(a) with a $R^2 = 0.999$. The calculated effective nucleation rate $\Gamma_{\rm eff} \approx (0.03)^2 t^2 - 0.3 t + 0.2 $ in panel (b) is used in panel (a) to plot the Avrami equation. }
\end{figure}

\section{KJMA framework}

The KJMA theory is a standard stochastic of nucleation and growth, the model describes how liquids transform to solid at constant temperature.\cite{Avrami1939} This model describes the kinetics of crystallization, in general it can be applied to other changes of phase.  This model predicts a smooth change of the order parameter that defines the phase like a compressed exponential form $\mu (t) = 1 -  \exp[-( \Gamma \, t)^2] $, the so-called Avrami equation for a 1D system, where $\Gamma$ is the nucleation rate. Here $\mu$ is the fraction of occupied sites.  The KJMA frame work allows to quantitatively understanding of the out-equilibrium dynamics introduced the previous section. 

In general, we can expect the fraction $\mu$ follows\cite{Avrami1939,Kol1937} 
\begin{equation}
	\mu(t) = 1 - \exp\left[- \Gamma^2 \int_0^t R(t-\tau)  d\tau  \right], 
\end{equation}

where $R(t)$ is the growth rate. For a lattice of $L$ sites the quantum analog of the fraction $\mu$ is equal to the density of particles, $\mu(t) = \bar{n}(t) = \sum_k \langle n_k(t) \rangle / L$.  The effective nucleation rate $\Gamma_{\rm eff} (t) = \Gamma^2 \int_0^t R(t-\tau)  d\tau$ can be calculated from the density $\bar{n}(t)$, and therefore the characteristics of the nucleation process.  In the weak coupling regime to the environment the structure of the growth rate is not trivial. While the lattice is populated coherent processes becomes relevant until the number of neighbor particles is enough to destroy it. This effect is not observed from the density of particles, see Fig. \ref{fig:plot1}(a),  but it is more evident from the behavior of the effective nucleation rate, see Fig. \ref{fig:plot1}(b). Once the lattice start to be occupied from incoming particles the density $\bar{n}(t)$ is well described by the Avrami equation, i.e. at the beginning the effective growth $\Gamma_{\rm eff} \propto t^2$, and $\Gamma^2 R(t) \propto t$, see Figs. \ref{fig:plot1}(b) and \ref{fig:plot1}(d). There we have delimited the region in the time evolution where the the system can be modeled by the Avrami equation with a trivial growth rate.          

\begin{figure}[h]
	\centering
	\includegraphics[width=\textwidth]{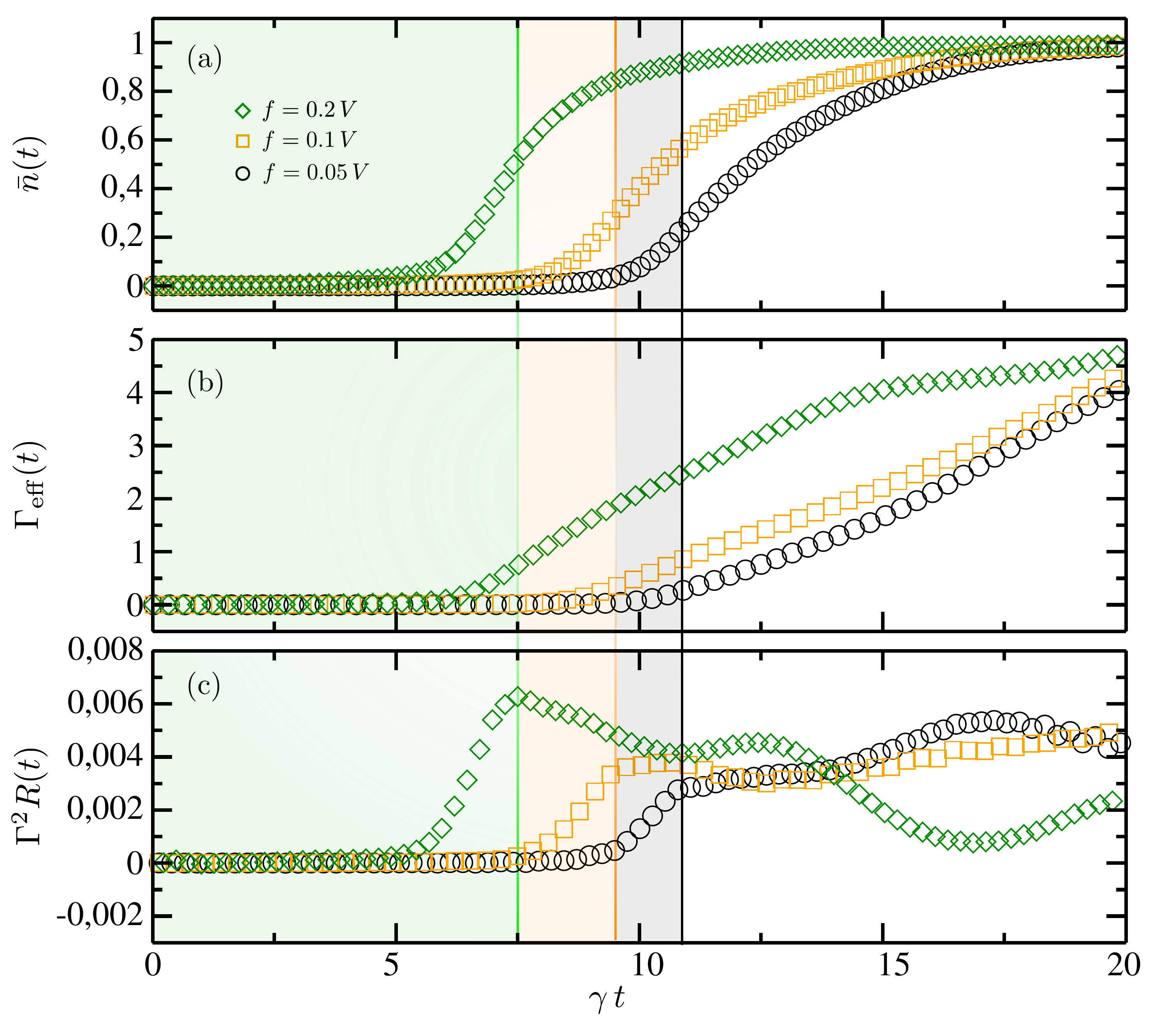}
	\caption{\label{fig:plot2} (a) Time evolution of the particle density for a lattice of $L = 200$ sites for different values for the modulation strength in units of the interaction energy $V = 1$. The damping rate used is $\gamma =  1 \times 10^{-2} V << f$. In addition the time evolution of the effective nucleation rate is depicted in (b). In panel (c) we show the behavior of the growth rate function. }
\end{figure}

\subsection{Moderated driving strength}
For a moderated driving strength the effective rate $\Gamma_{\rm eff}$ exhibit a very rich behavior, since the modulation strength is not strong enough to drive the system to a classical regime, but it can induced coherent processes. In the first stage of the population process, see the shadowed regions in Figure \ref{fig:plot2}, the nucleation rate $\Gamma$, which is proportional to the line curvature, is larger as stronger is the driving amplitude $f$. It is depicted in the particle density behavior in Figure \ref{fig:plot2}(a), there the density approaches to its maximum value faster for higher values of the driving strength. The shallow regions in Figure \ref{fig:plot2} are limited by the linear behavior of the growth rate (see Figure \ref{fig:plot2}(c)), before that region limit we expect the nucleation process is trivial and $R \propto t$. After that region the particle density reaches the maximum value following a non-trivial growing law ($R(t) \not\propto t $),  this is very interesting since the driving strength is not strong enough for consider a classical regime ($f \ll  V$) but still strong to overcome dissipative effect from the external environment.

\section{Conclusions}

We have consider the nucleation and growth process of neutral atoms in an one dimensional out-equilibrium optical lattice. The lattice is driven out -of equilibrium by the an external modulation.  The population in the lattice is described by the KJMA model, which explains  crystallization processes in liquids and other classical setups. The model describes the population growth until coherence states emerges from the inter-particle interaction. The effective nucleation rate and the growth rate can be calculated from the particle density measurement. This allows to connect classical nucleation and growth with the quantum dynamics in optical lattices. As a perspective of this work could be the connection of the growth rate with the space and time correlation function in the number of particle in the lattice.     

\ack
   
V. L. was supported by Universidad Santiago de Cali Project No 935-621115-N23, and M.C. was supported by Colciencias Project No 123365842816 FP44842-014-2015.

\end{document}